
\def\GG{\langle\alpha_sG^2\rangle}
\def\gev#1{ {\rm GeV^{#1}} }
\def\lvac{\langle\Omega\vert}
\def\rvac{\vert\Omega\rangle}

\def\frac#1#2{ { {#1}\over{#2} } }

\def\pder#1{ { {\partial}\over{\partial #1}} }
\def\G2{\langle G^2\rangle}
\def\alpi{ { {\alpha_s}\over{\pi} } }
\def\G3{\langle gG^3\rangle}

\font\blrm=cmbx10 scaled\magstep1
\magnification=\magstep1
\pageno=0
\baselineskip=20pt
\footline={\ifnum\pageno=0 \hfil \else \hss\folio\hss \fi}

\centerline{~~~}
\vskip1.5truecm
\centerline{{\blrm Mass of Pseudoscalar Gluonium: A Higher-Loop }}
\centerline{{\blrm QCD Sum-Rule Estimate}}
\vskip 1.5truecm
\centerline{{\bf D. Asner and R.B. Mann}}
\centerline{Department of Physics}
\centerline{University of Waterloo}
\centerline{Waterloo, Ontario, N2L 3G1 Canada}
\vskip 0.5truecm
\centerline{{\bf J.L. Murison and T.G. Steele\footnote\dag
{{\sl Address after July 1, 1992: Dept. of Physics, University of
Saskatchewan, Saskatoon, Saskatchewan, S7N 0W0 Canada}}
}}
\centerline{Department of Physics}
\centerline{University of Western Ontario}
\centerline{London, Ontario, N6A 3K7 Canada}
\vskip 1.0truecm
\noindent
{\bf Abstract:}
Higher-loop corrections to the pseudoscalar ($0^{-+}$) gluonium
correlation function
will be used to obtain the leading gluon condensate contributions to the
subtraction-independent QCD sum-rules.  The effect of these higher-loop
corrections on the sum-rule estimates of the pseudoscalar gluonium mass
will be investigated.  The final results of this analysis compare
favourably
with $SU(3)$ lattice simulations.

\vfill\eject

\noindent
The existence of QCD bound states consisting solely of gluons is one of the
conceptual predictions of QCD that has not been experimentally verified
[1].  Experimental searches for gluonia states are hampered by the
possibility of mixing with ordinary quark mesons of the same quantum
numbers, so a dynamical signature is necessary to identify the gluonium
content of observed mesons [2].

An approach to modelling mesons in QCD which has been successful in many
applications is QCD sum-rules [3].  In this approach the condensates of QCD
are included in the correlation functions of currents, providing a
parameterization of non-perturbative vacuum effects.

Although sum-rules have been useful for quark mesons, it has been difficult
to obtain a conclusive analysis of gluonium [4-7].  Part of this difficulty
for scalar gluonium
can be traced to a reliance on low-energy theorem  subtraction constants
to introduce the gluon condensate $\GG$  (the most reliably determined
gluonic condensate) into the sum-rule analysis.  In contrast, sum-rules
independent of the low-energy subtraction constants do not contain $\GG$ to
lowest order.  Thus higher-loop contributions may be significant
because they actually provide the leading $\GG$ behaviour in the
(subtraction-independent) sum-rules.

In previous work, the effect of these higher-loop corrections [8] was analyzed
for scalar ($0^{++}$) gluonium.
It was found that higher-loop effects were
significant in the subtraction-independent sum-rules,
leading to a mass prediction $m_{0^{++}}=(1.7 \pm 0.2)\gev{}$
for pure QCD [9], a result in reasonable agreement with lattice predictions
[10].

The purpose of this paper is to investigate higher-loop effects on the
sum-rule for pseudoscalar ($0^{-+}$) gluonium in pure QCD.  For the
pseudoscalar, the $N_f=0$ (pure QCD) limit is likely a better estimate of the
gluonium mass when quark effects are included than in the scalar case as
will be discussed below.  Our sum-rule estimate  of the pseudoscalar mass
and the $m_{0^{-+}}/m_{0^{++}}$ mass ratio will be compared with $SU(3)$
lattice values [10] and with previous sum-rule estimates [6,7].

The correlation function for pseudoscalar gluonium in pure QCD
is defined in terms of a renormalization group (RG) invariant current [11].
$$
\eqalign{
\Pi(q^2)&=i\int d^4x\,e^{iq\cdot x} \lvac T\left(j(x)j(0)\right)\rvac\cr
j(x)&=\alpha_sG^a_{\mu\nu}\tilde{G}^a_{\mu\nu}\quad ,\, \tilde{G}^a_{\mu\nu}
=\frac{1}{2} \epsilon_{\mu\nu\alpha\beta}G^a_{\alpha\beta}
}
\eqno(1)
$$
Perturbative contributions to $\Pi(Q^2=-q^2)$ are known to two-loops [12],
although some effort is required to relate these results to the following
expression.
$$
\eqalign{
\Pi(Q^2)&=-2\left( \alpi\right)^2Q^4\log{\frac{Q^2}{\nu^2}}\cr
&\quad\times\left(
1+\alpi\left[ \frac{31}{12}N_c-\frac{N_f}{6}
+\left(\frac{11}{12}N_c
+ \frac{N_f}{6}\right)\log{\frac{Q^2}{\nu^2}}
\right]\right)\cr
&+\,{\rm QCD~condensate~terms}
}
\eqno(2)
$$
Divergent constants and polynomials in $Q^2$ have been ignored since they
are zero when the sum-rule is formed.

The dependence on the number of
flavours $N_f$ and colours $N_c$ has been explicitly given in (2) to illustrate
that the flavour dependence is rather weak.  In particular, the transition
between pure QCD and $N_f=3$ is negligible.
$$
\eqalign{
\Pi(Q^2)&=-2\left( \alpi\right)^2Q^4\log{\frac{Q^2}{\nu^2}}\left[ 1+
\alpi\left(\frac{29}{4} -\frac{9}{4}\log{\frac{Q^2}{\nu^2}}\right)\right]\quad
N_c=3=N_f\cr
\Pi(Q^2)&=-2\left( \alpi\right)^2Q^4\log{\frac{Q^2}{\nu^2}}\left[ 1+
\alpi\left(\frac{31}{4}-\frac{11}{4}\log{\frac{Q^2}{\nu^2}}\right)\right]\quad
N_c=3 \quad N_f=0
}
\eqno(3)
$$
The situation is quite different for scalar gluonium, where apart from a
normalization constant $A$, the perturbative contributions to the
correlation function $\Phi(Q^2)$ are
$$
\eqalign{
\Phi(Q^2)&=A\left( \alpi\right)^2Q^4\log{\frac{Q^2}{\nu^2}}\left[ 1+
\alpi\left(\frac{41}{4} -\frac{9}{4}\log{\frac{Q^2}{\nu^2}}\right)\right]\quad
N_c=3=N_f\cr
\Phi(Q^2)&=A\left( \alpi\right)^2Q^4\log{\frac{Q^2}{\nu^2}}\left[ 1+
\alpi\left(\frac{51}{4} -\frac{11}{4}\log{\frac{Q^2}{\nu^2}}\right)\right]\quad
N_c=3 \quad N_f=0
}
\eqno(4)
$$
illustrating a stronger flavour dependence.  The weak flavour dependence in
the pseudoscalar case, combined with an estimated small mixing  [13] with
quark mesons,
suggest that the pure QCD prediction of the pseudoscalar mass is a good
approximation when quark effects are included.

The gluon condensate $\GG$ contributions to the correlation function of the
pseudoscalar current have only been
calculated to lowest order [4].  The general form of the $\GG$ portion of
$\Pi(Q^2)$ to one-loop, with the dependence on momentum given explicitly,
is
$$
\eqalign{
\Pi(Q^2)&=\alpi\left[ b_0+\alpi\left( b_1\log{\frac{Q^2}{\nu^2}}+
b_1^\prime\right)\right]\GG+\ldots\cr
b_0&=-4\pi;\quad b_1,\,b_1^\prime{\rm ~unknown}\qquad .
}
\eqno(5)
$$
The terms proportional to $b_0$, $b_1^\prime$ do not contribute to the sum-rule
since they are independent of the momentum $Q^2$.  Thus the leading $\GG$
contribution to the sum-rule comes from the one-loop logarithmic correction
proportional to $b_1$.

The one-loop calculation was carried out explicitly
for scalar gluonium to demonstrate that the operator-product expansion
coefficient is infrared finite [8].  However, an explicit calculation is
not necessary if the constant $b_1$ is only needed for the purpose of a
sum-rule analysis.  Since $\Pi(Q^2)$ is constructed from an RG invariant
current, the renormalized correlation function satisfies the following RG
equation
$$
\eqalign{
0&=\left(\nu\pder{\nu} +\beta\alpha_s\pder{\alpha_s}+{\rm
condensate~anomalous~dimensions}\right)\Pi(Q^2)\cr
\beta&=\beta_1\alpi=\left(-\frac{11}{6}N_c+\frac{N_f}{3}\right)\alpi  \quad
. } \eqno(6)
$$
For the gluon condensate terms, the anomalous dimension of
$\GG$ is zero to  one-loop.  Applying (6) to the $\GG$ dependence in (5)
shows that RG  invariance determines $b_1$.
$$
\eqalign{ 0&=-
2b_1\left(\alpi\right)^2\GG+b_0\beta_1\left(\alpi\right)^2\GG+ {\rm
terms~independent~of~}\GG\cr
b_1&=\frac{1}{2}\beta_1b_0 =-
\frac{11}{4}b_0 \,\,(N_c=3,\,N_f=0) } \eqno(7)
$$

Contributions to $\Pi(Q^2)$ from dimension six and dimension eight gluonic
condensates are also known to lowest order [4].  As will be seen below, the
one-loop logarithmic correction to $\langle gG^3\rangle=\langle g
f_{abc}G^a_{\mu\nu}G^b_{\nu\rho}G^c_{\rho\mu}\rangle$ is the leading
contribution in one of the sum-rules and is thus important.  Writing the
$\langle gG^3\rangle$ terms as
$$
\Pi(Q^2)=c_0\left(\alpi\right)^2\frac{\langle gG^3\rangle}{Q^2}+
c_1\left(\alpi\right)^3\log{\frac{Q^2}{\nu^2}}\frac{\langle gG^3\rangle}{Q^2}
\eqno(8)
$$
and applying the RG equation including the $\G3$ anomalous dimension [14],
$$
0=\left( \nu\pder{\nu}+\beta\alpha_s\pder{\alpha_s}+\frac{7}{11}\G3
\pder{\G3}\right)\Pi(Q^2)
\eqno(9)
$$
leads to the result
$$
c_1=-\frac{29}{4} c_0 \quad (N_c=3,\, N_f=0)
\eqno(10)
$$

Collecting the above results leads to the following expression for
$\Pi(Q^2)$, valid to two-loops in perturbative terms, one-loop in $\GG$ and
$\G3$, and to lowest order in higher dimension condensates.
$$
\eqalign{
\Pi(Q^2)&=Q^4\log{\frac{Q^2}{\nu^2}}\left[a_0+a_1\log{\frac{Q^2}{\nu^2}}
\right]\cr
&+\alpi\left(b_0+b_1^\prime\alpi\right)\GG+
\left(\alpi\right)^2b_1\log{\frac{Q^2}{\nu^2}}\,\GG\cr
&+\left(\alpi\right)^2\left[ c_0+\alpi
c_1\log{\frac{Q^2}{\nu^2}}\right]\frac{\G3}{Q^2}+d_0\frac{\langle \alpha^2_s
G^4\rangle}{Q^4}
\cr
a_0&=-2\left(\alpi\right)^2\left[1+\frac{31}{4}\alpi\right]\quad
a_1=\frac{11}{2}\left(\alpi\right)^3
\cr
b_0&=-4\pi\quad b_1=-\frac{11}{4}b_0=11\pi\cr
c_0&=-8\pi^2\quad c_1=-\frac{29}{4}c_0=8\pi^2\frac{29}{4}\cr
d_0&=8\pi^2\alpi\quad \langle \alpha^2_sG^4\rangle =\alpha^2_s\left(
\langle \left(f_{abc}G^b_{\mu\nu}\right)^2\rangle
+10\langle\left(f_{abc}G^b_{\mu\rho}G^c_{\rho\nu}\right)^2\rangle
\right)
}
\eqno(11)
$$

The correlation function satisfies a dispersion relation with two
subtraction constants, relating the QCD prediction $\Pi(Q^2)$ to the
phenomenological quantity ${\rm Im}\Pi(t)$:
$$
\Pi(Q^2)=\Pi(0)-\Pi^\prime(0)Q^2+\frac{Q^4}{\pi}\int^\infty_0 dt\,
\frac{{\rm Im}\Pi(t)}{t^2\left(t+Q^2\right)} \quad .
\eqno(12)
$$
The subtraction constant $\Pi^\prime(0)$ is the slope of the $U_A(1)$
topological charge, and has been estimated using sum-rule techniques [6].
Families of sum-rules can be constructed from
\vfill\eject
\noindent
(12) by
Borel-transforming the dispersion relation weighted with (positive) powers
of $Q^2$ [3].
$$
\eqalign{
{\cal R}_k(\tau, s_0)&=\frac{1}{\tau}\hat{L}\left[ Q^{2k}\Pi(Q^2)\right]
-\frac{1}{\pi}\int_{s_0}^\infty dt\, t^ke^{-t\tau} {\rm Im}\Pi^{{\rm eqn.
11}}(t)\cr
&=\frac{1}{\pi}\int_0^{s_0}dt\, t^ke^{-t\tau}{\rm Im}\Pi(t)\cr
\hat{L}&\equiv \lim_{{\scriptstyle N\rightarrow \infty \atop \scriptstyle
Q^2\rightarrow\infty}\atop \scriptstyle N/Q^2=\tau}
\frac{\left(-Q^2\right)^N}{(N-1)!}\left(\pder{Q^2}\right)^N
}
\eqno(13)
$$
The scale $\sqrt{s_0}$ is the continuum threshold representing duality
between resonance physics and QCD and $\hat{L}$ is the Borel transform
operator.  The continuum threshold is constrained by the finite-energy
sum-rule [15].
$$
{\cal F}_0(s_0)=\frac{1}{\pi}\int_0^{s_0} dt\,{\rm Im}\Pi(t)
\eqno(14)
$$
After some calculation, the sum-rules of interest are obtained from the
QCD correlation function $\Pi(Q^2)$ in (11).
$$
\eqalign{
{\cal R}_0(\tau, s_0)&=
-\frac{2a_0}{\tau^3}\left[1-\rho_2(s_0\tau)\right]\cr
&-\frac{4a_1}{\tau^3}\left[
\frac{3}{2}-\gamma_E-\rho_2(s_0\tau)\log{s_0\tau}-\frac{3}{2}e^{-s_0\tau}\left(
1+\frac{1}{3}s_0\tau\right)-E_1(s_0\tau)\right]\cr
&-\left(\alpi\right)^2\frac{b_1}{\tau}\left(1-e^{-s_0\tau}\right)\GG
+c_0\left(\alpi\right)^2\G3 +\cr
&-c_1\left(\alpi\right)^3\G3\left[\gamma_E+E_1(s_0\tau)\right]  +
d_0\tau\langle\alpha^2_sG^4\rangle
}
\eqno(15a)
$$
$$
\eqalign{
{\cal R}_1(\tau,s_0)&=-\frac{6a_0}{\tau^4}\left[1-\rho_3(s_0\tau)\right]\cr
&-\frac{12a_1}{\tau^4}\left[
\frac{11}{6}-\gamma_E-\rho_3(s_0\tau)\log{s_0\tau}-
\frac{3}{2}e^{-s_0\tau}
\left(1+\frac{1}{3}s_0\tau\right)-\frac{1}{3}\rho_2(s_0\tau)\right.\cr
&\qquad\qquad\qquad -\biggl. E_1(s_0\tau)
\biggr]\cr
&-\frac{b_1}{\tau^2}\left(\alpi\right)^2\left[1-\rho_1(s_0\tau)\right]\GG
\cr& +\frac{c_1}{\tau}
\left(\alpi\right)^3\left[1-e^{-s_0\tau}\right]\G3
-d_0\langle\alpha^2_sG^4\rangle
}
\eqno(15b)
$$
$$
\eqalignno{
{\cal F}_0(s_0)&=-\frac{1}{3}a_0s^3_0+\frac{2}{9} a_1s^3_0-b_1s_0\left(\alpi
\right)^2\GG
+c_0\G3&(16)\cr
\rho_k(x)&\equiv e^{-x}\sum_{j=0}^{k} \frac{x^j}{j!}\quad
\gamma_E\equiv{\rm Euler's~Constant}\approx 0.5772\cr
E_1(x)&\equiv \int_x^\infty dy\,\frac{e^{-y}}{y}\quad ({\rm Exponential~ Integral)
}}
$$
Renormalization group improvement of the sum-rules implies that the
running coupling constants
$\overline{\alpha}_s(1/\tau),\,\overline{\alpha}_s(s_0)$ respectively
appear in (15) and (16).  In principle, the condensates other than $\GG$
should also be RG improved (only $\GG$ is an RG invariant to one-loop).
These effects can be explicitly considered for $\G3$, and are found to
have a small effect, while for $\langle \alpha^2_s G^4\rangle$ the
vacuum saturization hypothesis used to estimate its numerical value [4,16]
leads to the (RG invariant) $\GG$ condensate.

As mentioned previously, notice that the leading $\GG$ behaviour in both
sum rules, and the leading $\G3$ correction in the ${\cal R}_1$
sum-rule comes from the one-loop logarithmic terms in $\Pi(Q^2)$.  These
contributions have not been considered in previous studies of
pseudoscalar gluonium, motivating our analysis of the sum-rule mass
estimates.

The sum-rules ${\cal R}_k(\tau,s_0)$ relate a QCD prediction to a
phenomenological model for ${\rm Im}\,\Pi(t)$.  The simplest model is the
narrow-width approximation for the lightest resonance,
$$
{\rm Im}\,\Pi(t)=\pi f^2\delta(t-M^2)
\eqno(17)
$$
where $M$ is the mass of the $0^{-+}$ state and $f$ is its coupling to the
pseudoscalar current.  This model leads to the following family of
sum-rules relating QCD expressions to resonance properties of the state.
$$
\eqalign{
{\cal R}_0(\tau, s_0)&= f^2M^4e^{-M^2\tau}\cr
{\cal R}_1(\tau,s_0)&=f^2M^6e^{-M^2\tau}\cr
{\cal F}_0(s_0)&=f^2M^4 \cr
}
\eqno(18)
$$
Taking the ratio of ${\cal R}_1/{\cal R}_0$ it is easily observed that
$$
M^2=\frac{ {\cal R}_1(\tau,s_0)}{{\cal R}_0(\tau,s_0)}
\eqno(19)
$$
For small $\tau$ (high-energy) the dominance of the lightest resonance in
the phenomenological model fails, while for large $\tau$ (low-energy) the
OPE is no longer a good approximation.  Thus at intermediate values of
$\tau$ there should be a slowly varying region of the ratio (19).

To obtain a QCD prediction of the resonance properties, the conventional
($\tau$ stability) procedure [3] is to fix $s_0$ and then identify the
$\tau$ stationary point of ${\cal R}_1/{\cal R}_0$ as the ($s_0$-dependent)
value of $M^2$.  This value is then used to extract $f^2$ by locating the
stationary point of $e^{M^2\tau}{\cal R}_0/M^4$.  This procedure then
leads to $M$ and $f$ as functions of the continuum threshold $s_0$.  To
determine the final QCD prediction, the finite-energy sum-rule (FESR)
is used to
constrain $s_0$ by demanding maximum agreement with the lowest FESR [15].
$$
f^2(s_0)M^4(s_0)={\cal F}_0(s_0)
\eqno(20)
$$

The outcome of this procedure will obviously depend on the QCD parameters
used as input into the sum-rules.  In an exhaustive analysis of several
channels
it has been concluded that an acceptable range of the gluon condensate is
$\GG=3(0.05\pm 0.015)/\pi\,\gev{4}$ [17].  The dilute instanton gas
approximation $\G3\approx (0.27\,\gev{2})\GG$  [18] and vacuum
saturization $\langle \alpha^2_sG^4\rangle= \frac{15}{16}(\GG)^2$ [4,16]
will be
used for the higher-dimension gluonic condensates.  These values for the
higher-dimension condensates are in agreement with the conclusions for
the scalar glueball sum-rule [9].  Finally, $\Lambda_{\overline{MS}}$
will be allowed to vary over a generous range.  To summarize, the
following QCD parameter space will be considered.
$$
\eqalign{
\GG&= \frac{3}{\pi}(0.050\pm 0.015)\,\gev{4}\cr
\G3&=(0.27\,\gev{2})\GG\quad\langle \alpha^2_sG^4\rangle= \frac{15}{16}(\GG)^2
\cr
\overline{\alpha}_s(1/\tau)&=-\frac{4\pi}{11\log{
\tau\Lambda^2_{\overline{MS} } } }
\quad \Lambda_{\overline{MS}}=(0.15 \pm 0.05)\gev{}
}
\eqno(21)
$$
Two-loop corrections to the running coupling constant  will be
applied in the (two-loop) perturbative terms.

The results of the previously described algorithm for analyzing the
pseudoscalar
sum-rule are more conclusive than in the scalar channel.  It is possible
to find a value of $s_0$ leading to precise agreement\footnote\dag{
As $s_0$ is increased, the quantity ${\cal F}_0(s_0)-f^2(s_0)M^4(s_0)$
changes sign, and hence there exists $s_0$ in exact agreement with the
FESR.}
with the FESR
constraint, and also providing wide flat regions in the $\tau$ plots.
These plots are shown for the optimum $s_0$ and various choices of
parameter space in Figures 1-3, and summarized in Table 1.  As is
necessary in a sum-rule analysis, the $\tau$ stationary points occur at
intermediate values, so that convergence of the QCD sum rule (small
$\tau$) is balanced against dominance of the lightest resonance (large
$\tau$).

Table 1 exhibits the complete range of mass predictions resulting from
the QCD parameter space (21).  Thus over the range of parameters
considered, the $\tau$ stability analysis of the sum-rules satisfies the
FESR constraint, leading to the predictions $f=(0.30\pm0.05)\gev{}$,
$m_{0^{-+}}=(2.3\pm 0.2)\gev{}$, with the errors reflecting only the
parameter space uncertainties.

Our analysis leads to a heavier $0^{-+}$ mass prediction than in previous
sum rule estimates [6,7], indicating that the higher-loop corrections
considered here are indeed important.  However, our results are in
reasonable agreement with $SU(3)$ lattice simulations which find
$m_{0^{-+}}=(6.3\pm 0.7)\sqrt{K}\approx (2.8 \pm 0.3)\gev{}$ [10].  For
mass ratios, using the (higher-loop) sum rule estimate for the scalar
glueball [9], we find $m_{0^{-+}}/m_{0^{++}}=(1.4\pm 0.3)$ in good
agreement with the $SU(3)$ lattice ratio $m_{0^{-+}}/m_{0^{++}}=(1.8\pm
0.3)$.

Since our results disagree with the other sum-rule estimates, we feel that it
is incumbent upon us to check our calculations using the same corrections
as in [6].  A value of the continuum threshold can be found satisfying the
sum-rule constraint, and the corresponding plots are shown in Figure 4.  As
is evident a mass scale of about $1.7\gev{}$ results, in agreement with
[6].  It can thus be concluded that the neglected contribution of the
gluon condensate, providing the leading $\GG$ behaviour in the sum-rule,
has a significant effect on the sum-rule analysis.

In conclusion, the effect of higher-loop corrections on the sum-rule
predictions of the pseudoscalar gluonium mass have been investigated in
pure QCD.  These higher-loop corrections provide the leading $\GG$
behaviour in the (subtraction-independent) sum-rules, and thus their
effects are significant.  In the sum-rule analysis, a continuum threshold
can be found in precise agreement with the lowest FESR, leading to
optimum predictions $m_{0^{-+}}=(2.3\pm 0.2)\gev{}$ over a wide range of
QCD parameter space.  These results compare well with lattice values,
both for the pseudoscalar mass, and the pseudoscalar-scalar mass ratio.

\noindent
{\blrm Acknowledgements}\hfil\break
Many thanks to E. Bagan for collaborative work leading to this project.
We are grateful for the financial support of the Natural Sciences and
Engineering Research Council of Canada.

\beginsection{{\blrm \centerline{References}}}

\parindent=0pt
\smallskip
\item{{\bf [1]}} Particle Data Group, Phys. Lett. {\bf 239B} (1990).
\smallskip
\item{{\bf [2]}} M.S. Chanowitz, Phys. Lett. {\bf 187B} (1987) 409.
\smallskip
\item{{\bf [3]}} M.A. Shifman, A.I. Vainshtein, V.I. Zakharov, Nucl. Phys. {\bf
B147} (1979) 385, 448; \hfil\break
 L.J. Reinders, H. Rubenstein, S. Yazaki, Phys. Rep. {\bf C127} (1985) 1.
\smallskip
\item{{\bf [4]}} V.A. Novikov, M.A. Shifman, A.I. Vainshtein, V.I.
Zakharov, Phys. Lett. {\bf 86B} (1979) 347.
\smallskip
\item{{\bf [5]}} M.A. Shifman, Z. Phys. {\bf C9} (1981) 347;
\hfil\break
S. Narison, Phys. Lett. {\bf 125B} (1983) 501;
\hfil\break
S. Narison, Z. Phys. {C26} (1984) 209;
\hfil\break
J. Bordes, V. Gim\`enez, J.A. Pe\~narrocha, Phys. Lett. {\bf 223B} (1989) 251.
\smallskip
\item{{\bf [6]}} K. Senba, M. Tanimoto, Phys. Lett. {\bf 105B} (1981) 297.
\smallskip
\item{{\bf [7]}} S. Narison, Phys. Lett. {\bf 255B} (1991) 101.
\smallskip
\item{{\bf [8]}} E. Bagan, T.G. Steele, Phys. Lett. {\bf 234B} (1990) 135.
\smallskip
\item{{\bf [9]}} E. Bagan, T.G. Steele, Phys. Lett. {\bf 243B} (1990) 413.
\smallskip
\item{{\bf [10]}} C. Michael, M. Teper, Phys. Lett. {\bf 206B} (1988) 299;
\hfil\break
C. Michael, M. Teper, Nucl. Phys. {\bf B314} (1989) 347.
\smallskip
\item{{\bf [11]}} D. Espriu, R. Tarrach, Z. Phys. {\bf C16} (1982) 77.
\smallskip
\item{{\bf [12]}} A.L. Kataev, N.V. Krasnikov, A.A. Pivovarov, Nucl. Phys.
{\bf B198} (1982) 508.
\smallskip
\item{{\bf [13]}} S. Narison, N. Pak, N. Paver, Phys. Lett. {\bf 147B}
(1984) 162.
\smallskip
\item{{\bf [14]}} S. Narison, R. Tarrach, Phys. Lett. {\bf 125B} (1983)
217.
\smallskip
\item{{\bf [15]}} R.A. Bertlmann, G. Launer, E. de Rafael, Nucl. Phys. {\bf
B250} (1985) 61.
\smallskip
\item{{\bf [16]}} E. Bagan, J.I. Latorre, P. Pascual, R. Tarrach, Nucl.
Phys. {\bf B254} (1985) 555.
\smallskip
\item {{\bf [17]}} V. Gim\`enez, J. Bordes, J.A. Pe\~narrocha,
Phys. Lett. {\bf 223B} (1989) 251.
\smallskip
\item{{\bf [18]}} V.A. Novikov, M.A. Shifman, A.I. Vainshtein, V.I.
Zakharov, Nucl. Phys. {\bf B191} (1981) 301.
\smallskip

\beginsection{{\blrm \centerline{Figure Captions}}}

\noindent
Figures 1-3 illustrate the sum-rule estimates of $f$ and $M$ for the
optimum values of $s_0$.\hfil\break
{\bf Fig. 1:~}$\GG=0.033\,\gev{4},~\Lambda_{\overline{MS}}=0.2\,\gev{}~,
\sqrt{s_0}=2.7\,\gev{}$\hfil\break
{\bf Fig. 2:~}$ \GG=0.048\,\gev{4},~\Lambda_{\overline{MS}}=0.15\,\gev{}~,
\sqrt{s_0}=2.9\,\gev{}$\hfil\break
{\bf Fig. 3:~}$\GG=0.062\,\gev{4},~\Lambda_{\overline{MS}}=0.1\,\gev{}~,
\sqrt{s_0}=3.1\,\gev{}$\hfil\break\hfil\break

\noindent
{\bf Fig. 4:~} Illustration of the sum-rule analysis neglecting the
higher-loop effects.  The parameters are:
$\GG=0.04\,\gev{4},~\Lambda_{\overline{MS}}=0.15\,\gev{}~,
\sqrt{s_0}=2.2\,\gev{}$ and lead to an agreement with the FESR of
$({\cal F}_0-f^2M^4)/{\cal F}_0=1 \times 10^{-2}$.

\bye